\documentclass[12pt]{article}
\usepackage{graphicx}
\usepackage[numbers]{natbib}
\usepackage{rotating}

\newcommand\pubnumber{}
\newcommand\pubdate{\today}

\def\institute{Department of Physics and Astronomy\\
Michigan State University, East Lansing, MI 48824, USA}

\def\Title#1{\begin{center} {\Large #1 } \end{center}}
\def\Author#1{\begin{center}{ \sc #1} \end{center}}
\def\Address#1{\begin{center}{ \it #1} \end{center}}

\newcommand\pubblock{\rightline{\begin{tabular}{l} \pubnumber\\
         \pubdate  \end{tabular}}}
\newenvironment{Abstract}{\begin{quotation}  }{\end{quotation}}
\newenvironment{Presented}{\begin{quotation} \begin{center} 
             PRESENTED AT\end{center}\bigskip 
      \begin{center}\begin{large}}{\end{large}\end{center} \end{quotation}}
\def\Acknowledgements{\bigskip  \bigskip \begin{center} \begin{large}
             \bf ACKNOWLEDGEMENTS \end{large}\end{center}}

\begin{document}
\begin{titlepage}
\pubblock

\vfill
\Title{Selected Topics from Top Mass Measurements at the Tevatron}
\vfill
\Author{ Reinhard Schwienhorst,\\  on behalf of the CDF and D0 Collaborations}
\Address{\institute}
\vfill
\begin{Abstract}
The most recent results of the top-quark mass measurements at the Tevatron at Fermilab are presented. Data were collected in proton-antiproton collisions at $\sqrt{s}=1.96$~TeV by the CDF and D0 experiments. Top quark mass measurements in the lepton+jets, dilepton and alljet final states as well as their combination and the extraction of the mass from the cross-section measurement are presented.
\end{Abstract}
\vfill
\begin{Presented}
$9^{th}$ International Workshop on Top Quark Physics\\
Olomouc, Czech Republic,  September 19--23, 2016
\end{Presented}
\vfill
\end{titlepage}
\def\thefootnote{\fnsymbol{footnote}}
\setcounter{footnote}{0}

\section{Introduction}

The top quark was discovered in 1995 at the Tevatron at
Fermilab~\cite{Abe:1995hr,Abachi:1995iq}. It is the heaviest fundamental particle and has
a short lifetime, decaying before hadronization can take place. The top quark thus plays
a crucial role in studying a bare quark. In addition, its large Yukawa coupling implies that
the top quark may play a crucial role in electroweak symmetry breaking.

The mass of the top quark ($m_t$) is a fundamental parameter of the standard model (SM) of
particle physics. Loop-corrections to the Higgs boson mass connect the top quark mass, the
$W$~boson mass and the Higgs boson mass.
A precise measurement of the top quark mass also determines whether the electroweak
vacuum is unstable, metastable, or stable~\cite{Degrassi:2012ry}.

The top quark mass measurements presented here are based on a dataset of approximately
10~fb$^{-1}$, collected in Run~II (2001 to 2011) at the Tevatron proton-antiproton collider
at a center-of-mass energy of $\sqrt{s}=1.96$~TeV. The top quark decays into a $W$~boson
and a $b$-quark almost 100\% of the time, and the top-quark pair final state is 
characterized by the decays of the two $W$~bosons into the lepton+jets, dilepton, and all-jets final state.

Two different top-quark mass measurements are presented: An extraction of the MC mass from the reconstruction of the top quark final state decay products, and an extraction of the pole mass from a measurement of the top quark pair production differential cross-section.
The top quark mass measurements at the Tevatron have large systematic uncertainties from
both theoretical modeling and experimental sources. For the MC mass measurement, the dominant source of systematic uncertainty is the jet energy scale (JES), which is constrained in the fit to data in some of the measurements. In addition, the measured mass relies on the implementation

\section{Top Quark MC Mass Measurements}
\label{sec:meas}

The two most common techniques for measuring the mass of the top quark are the template method
and the matrix element method. The template method interprets the distributions of a set of
variables that are sensitive to the top quark mass as probability densities. The matrix element (ME) technique explores the final-state topology of each event by calculating the probability of each event being signal for a given top quark mass hypothesis.

\subsection{Lepton+jets}
\label{sec:ljets}
The final top quark mass measurement in the lepton+jets final state by the D0
experiment~\cite{Abazov:2014dpa,Abazov:2015spa} utilizes the ME
method. Systematic uncertainties are reduced in the measurement through an updated detector calibration,
in particular improvements to the JES corrections for $b$-quark jets~\cite{Abazov:2013hda}. The jet energy scale factor is determined in situ in the fit to data. The top quark mass is measured with a precision of 0.43\%,
$m_t = 174.98 \pm 0.58$~(stat+JES)~$\pm 0.49$~(syst)~GeV. The largest sources of systematic uncertainty 
are the signal modeling and the residual JES, which account for 0.3~GeV and  0.21~GeV, respectively.

\subsection{All-jets}
\label{sec:aj}
The measurement in the all-jets final state by the CDF experiment~\cite{Aaltonen:2014sea} uses the
template method wit the full Run II dataset of 9.3~fb$^{-1}$. The event selection maximizes the signal fraction through the use of a neural network that is based on 13 kinematic variables. The decay products of the two
top quarks and the two $W$~bosons are reconstructed in a constrained kinematic fit. The top quark mass
is obtained together with the jet energy scale factor with an unbinned likelihood technique. The top quark mass
is measured with 1.1\% precision, 
$m_t = 175.07 \pm 1.19$~(stat)~$^{+1.55}_{-1.58}$~(syst)~GeV. The largest sources of systematic uncertainty
are from the residual JES and the trigger, each about 0.6~GeV.

The CDF measurement of the top quark mass in the $\;\slash{\!\!\!\!E}_T$ plus jets final state is sensitive to top
decays where one or more of the $W$~boson decay products is not reconstructed, including $\tau$~lepton
decays~\cite{Aaltonen:2013aqa}. Events are required to have large $\;\slash{\!\!\!\!E}_T$ and between four and six
jets, vetoing events with identified leptons. Similar to the all-jets analysis, the top quark mass is obtained from a fit to the reconstructed mass of the two top quarks and of the hadronically decaying $W$~boson. The top quark mass is measured with 1.1\% precision,
$m_t = 173.93 \pm 1.64$~(stat+JES)~$\pm 0.87$~(syst)~GeV. The largest sources of systematic uncertainty
are from the residual JES and the signal modeling, each about 0.4~GeV.

\subsection{Dilepton}
\label{sec:dl}
The measurement in the dilepton final state by the D0 experiment uses two methods to improve the sensitivity of the measurement to the mass value. The first method uses the neutrino weighting technique~\cite{D0:2015dxa}. For each event, a weight distribution as a function of the hypothetical top quark mass is obtained by integrating over the possible momenta for the two neutrinos, using the constraints from the top quark and $W$~boson masses. The mass is obtained from the width and mean of this weight distribution. The second method uses the matrix element technique~\cite{D0:2016ull}.  
The impact of the JES is reduced by propagating the the JES scale factor from the lepton+jets measurement~\cite{Abazov:2015spa} to both of these analyses. The results are then combined~\cite{D0dilcomb}.
The top quark mass is measured with 0.9\% precision,
$m_t = 173.50 \pm 1.31$~(stat)~$\pm 0.84$~(syst)~GeV. The dominant uncertainty is from the jet energy scale.

The CDF measurement in the dilepton final state uses a hybrid method in order to reduce the impact of the JES uncertainty~\cite{Aaltonen:2015hta}. The observable used in the template fit is a weighted average of two different mass estimators that have different sensitivity to the JES.  The top quark mass is measured with 1.9\% precision,
$m_t = 171.5 \pm 1.9$~(stat)~$\pm 2.5$~(syst)~GeV. The dominant uncertainty is from the jet energy scale.

\subsection{Tevatron combination}
\label{sec:comb}
\begin{figure}[!h!tbp]
  \centerline{\includegraphics[width=0.55\textwidth]{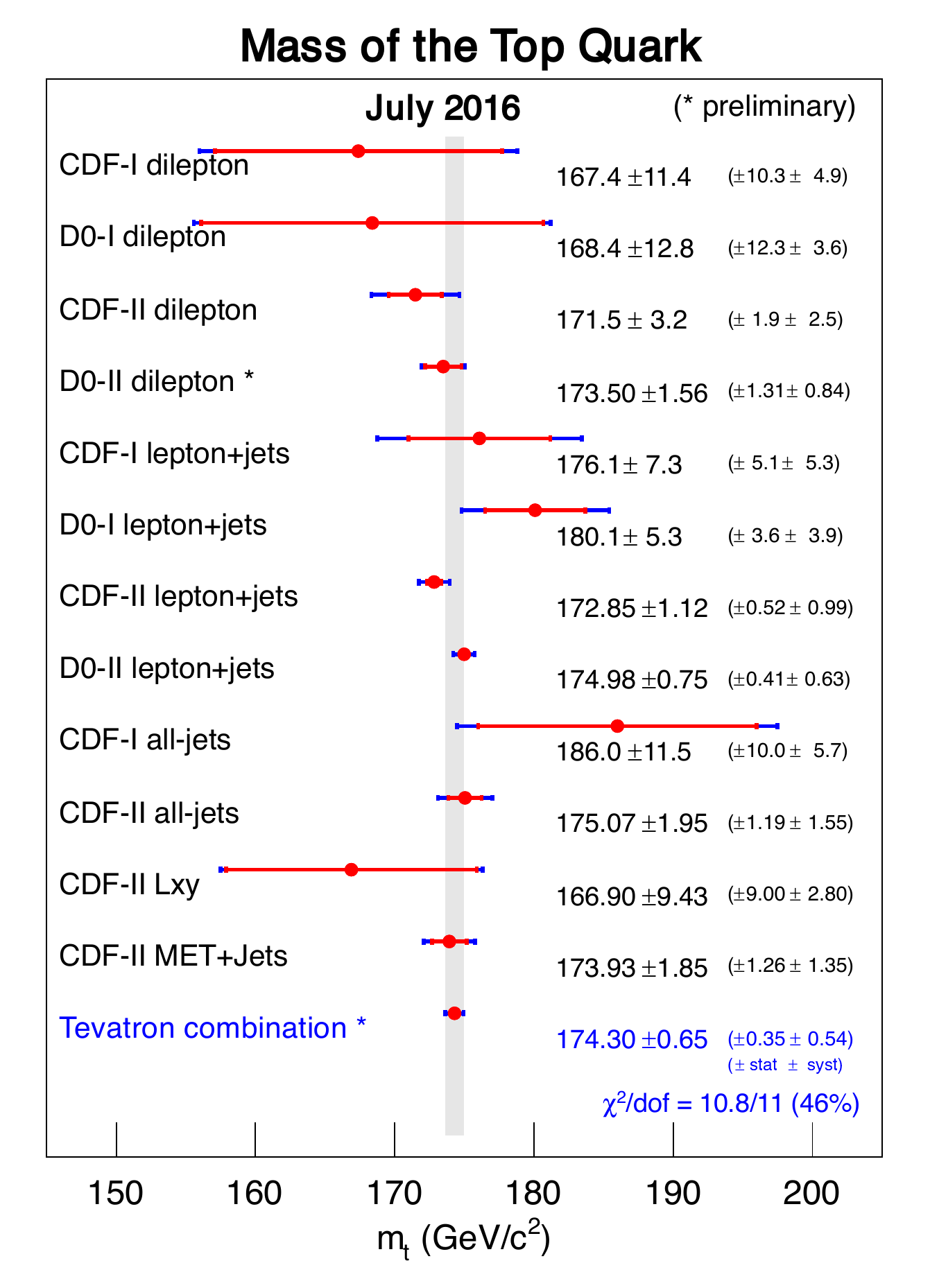}}
  \caption{Summary of Tevatron top quark mass measurements and their combination~\cite{Tevcomb}.
\label{fig:comb}
}
\end{figure}

The results described above and other measurements of the top quark mass at the Tevatron were combined in July 2016~\cite{Tevcomb}. A summary of the input measurements and the Tevatron combination is shown in Figure~\ref{fig:comb}. The  top quark mass is measured with 0.37\% precision,
$m_t = 174.34 \pm 0.37$~(stat)~$\pm 0.52$~(syst)~GeV. 

\section{Top pole mass extraction}
\label{Sec:pole}
The top quark pole mass is extracted from the total production cross-section~\cite{Abazov:2016ekt} and from the differential cross-section~\cite{D0pole}. The experimental measurements are compared to predictions at next-to-next-to leading order (NNLO)~\cite{Czakon:2016ckf}. The top-quark pole mass is measured with 1.5\% precision, $m_t = 169.1 \pm 2.5$~GeV. The dominant uncertainty is from the experimental determination of the differential distribution.

\begin{figure}[!h!tbp]
  \centerline{\includegraphics[width=0.7\textwidth]{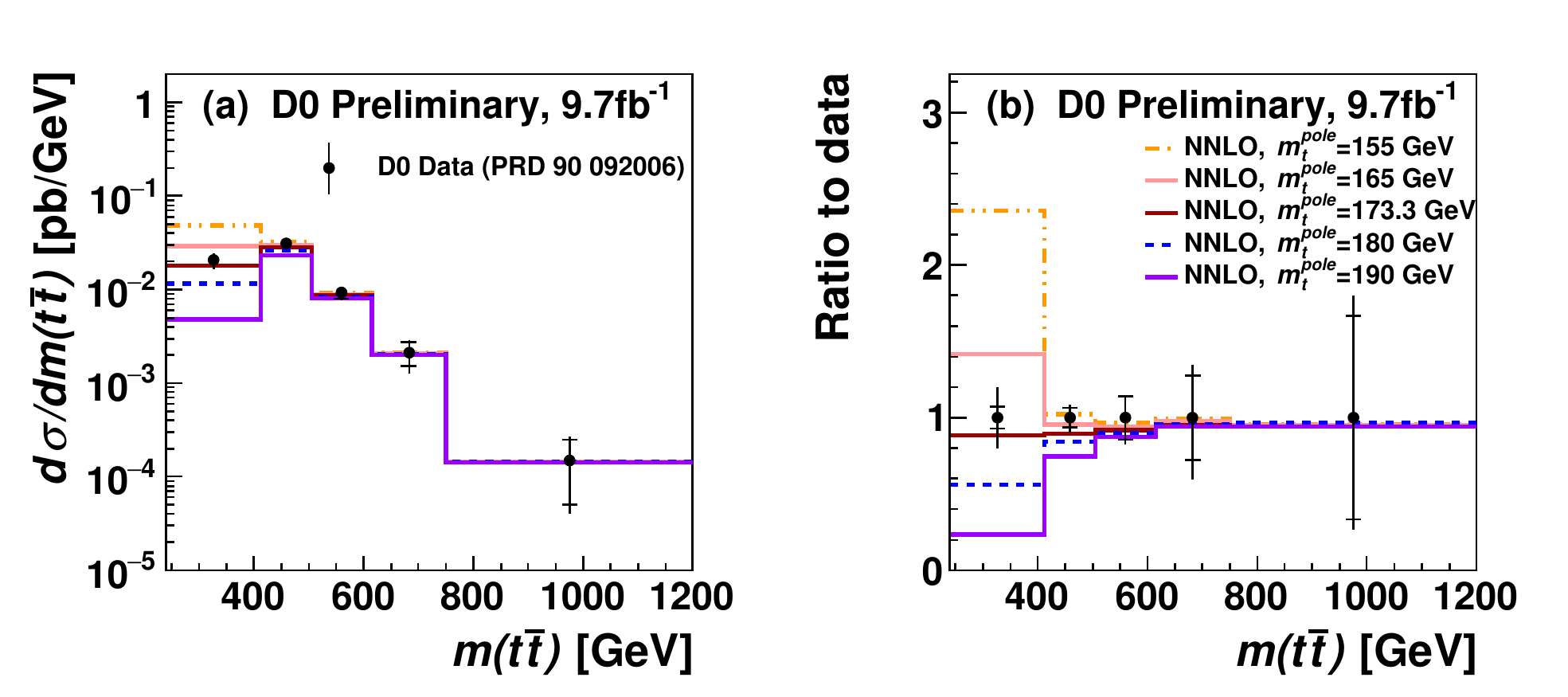}}
  \caption{Unfolded differential distribution of the invariant mass of the top-quark pair~\cite{D0pole}, comparing D0 data to NNLO predictions.
\label{fig:diff}
}
\end{figure}

\section{Conclusions}
\label{sec:concl}

Measurements of the top quark mass by the CDF and D0 experiments with the full Run~II dataset have been performed in top quark production and decay. These are among the worlds most precise measurements.

\Acknowledgements
This work was supported in part by the US National Science Foundation under grants
PHY-1068318 and PHY-1410972.


\nocite{*}
\bibliographystyle{aipnum-cp}%
\bibliography{Top2016TevTopMass}%

\end{document}